\documentclass[aps,prl,showpacs,twocolumn,lineno,groupedaddress]{revtex4}

\usepackage{graphicx}
\usepackage{epsfig}
\usepackage{dcolumn}
\usepackage{bm}
\usepackage{amssymb}   
\usepackage[usenames]{color} 

\newcommand{\beq}{\begin{equation}}
\newcommand{\eeq}{\end{equation}}
\newcommand{\barr}{\begin{eqnarray}}
\newcommand{\earr}{\end{eqnarray}}

\def\rperp{R_{\perp}}

\def\bq{\begin{quote}}
\def\eq{\end{quote}}

\def\spose#1{\hbox to 0pt{#1\hss}}
\def\lsim{\mathrel{\spose{\lower 3pt\hbox{$\mathchar"218$}}
 \raise 2.0pt\hbox{$\mathchar"13C$}}}
\def\gsim{\mathrel{\spose{\lower 3pt\hbox{$\mathchar"218$}}
 \raise 2.0pt\hbox{$\mathchar"13E$}}}

\def\bs{${B_s^0}$}

\def\bsdec{${B_s^0 \rightarrow J/\psi \phi}$}

\def\D0{D\O }

\def\GeVp{ {\ifmmode \;{{\mbox{\mathrm GeV}} / {\mbox\mathrm c}} \else
${{\mbox{\mathrm GeV}} / {\mbox\mathrm c}}$ \fi }}
\def\MeVp{ {\ifmmode \;{{\mbox{\mathrm MeV}} / {\mbox\mathrm c}} \else
${{\mbox{\mathrm MeV}} / {\mbox\mathrm c}}$ \fi }}
\def\MeV{ {\ifmmode \;{{\mbox{\mathrm MeV}} / {\mbox\mathrm c}^2} \else
${{\mbox{\mathrm MeV}} / {\mbox\mathrm c}^2}$ \fi }}
\def\GeV{ {\ifmmode \;{{\mbox{\mathrm GeV}} / {\mbox\mathrm c}^2} \else
${{\mbox{\mathrm GeV}} / {\mbox\mathrm c}^2}$ \fi }}

\providecommand{\tabularnewline}{\\}

\begin{document}


\hspace{5.2in}
\mbox{FERMILAB-PUB-07-007-E}

\title{Lifetime Difference and CP-Violating Phase in the $B_s^0$ System}
\date{January 9, 2007}

%
\author{                                                                      
V.M.~Abazov,$^{35}$                                                           
B.~Abbott,$^{75}$                                                             
M.~Abolins,$^{65}$                                                            
B.S.~Acharya,$^{28}$                                                          
M.~Adams,$^{51}$                                                              
T.~Adams,$^{49}$                                                              
E.~Aguilo,$^{5}$                                                              
S.H.~Ahn,$^{30}$                                                              
M.~Ahsan,$^{59}$                                                              
G.D.~Alexeev,$^{35}$
% ...
\\                                                                            
\vskip 0.30cm                                                                 
\centerline{(D\O\ Collaboration)}                                             
\vskip 0.30cm                                                                 
}                                                                             
\affiliation{                                                                 
\centerline{$^{1}$Universidad de Buenos Aires, Buenos Aires, Argentina}       
\centerline{$^{2}$LAFEX, Centro Brasileiro de Pesquisas F{\'\i}sicas,         
                  Rio de Janeiro, Brazil}                                     
\centerline{$^{3}$Universidade do Estado do Rio de Janeiro,                   
                  Rio de Janeiro, Brazil}                                     
\centerline{$^{4}$Instituto de F\'{\i}sica Te\'orica, Universidade            
                  Estadual Paulista, S\~ao Paulo, Brazil}                     
\centerline{$^{5}$University of Alberta, Edmonton, Alberta, Canada,           
                  Simon Fraser University, Burnaby, British Columbia, Canada,}
\centerline{York University, Toronto, Ontario, Canada, and                    
                  McGill University, Montreal, Quebec, Canada}                
\centerline{$^{6}$University of Science and Technology of China, Hefei,       
                  People's Republic of China}                                 
\centerline{$^{7}$Universidad de los Andes, Bogot\'{a}, Colombia}             
\centerline{$^{8}$Center for Particle Physics, Charles University,            
                  Prague, Czech Republic}                                     
\centerline{$^{9}$Czech Technical University, Prague, Czech Republic}         
\centerline{$^{10}$Center for Particle Physics, Institute of Physics,         
                   Academy of Sciences of the Czech Republic,                 
                   Prague, Czech Republic}                                    
\centerline{$^{11}$Universidad San Francisco de Quito, Quito, Ecuador}        
\centerline{$^{12}$Laboratoire de Physique Corpusculaire, IN2P3-CNRS,         
                   Universit\'e Blaise Pascal, Clermont-Ferrand, France}      
\centerline{$^{13}$Laboratoire de Physique Subatomique et de Cosmologie,      
                   IN2P3-CNRS, Universite de Grenoble 1, Grenoble, France}    
\centerline{$^{14}$CPPM, IN2P3-CNRS, Universit\'e de la M\'editerran\'ee,     
                   Marseille, France}                                         
\centerline{$^{15}$Laboratoire de l'Acc\'el\'erateur Lin\'eaire,              
                   IN2P3-CNRS et Universit\'e Paris-Sud, Orsay, France}       
\centerline{$^{16}$LPNHE, IN2P3-CNRS, Universit\'es Paris VI and VII,         
                   Paris, France}                                             
\centerline{$^{17}$DAPNIA/Service de Physique des Particules, CEA, Saclay,    
                   France}                                                    
\centerline{$^{18}$IPHC, IN2P3-CNRS, Universit\'e Louis Pasteur, Strasbourg,  
                   France, and Universit\'e de Haute Alsace,                  
                   Mulhouse, France}                                          
\centerline{$^{19}$IPNL, Universit\'e Lyon 1, CNRS/IN2P3, Villeurbanne, France
                   and Universit\'e de Lyon, Lyon, France}                    
\centerline{$^{20}$III. Physikalisches Institut A, RWTH Aachen,               
                   Aachen, Germany}                                           
\centerline{$^{21}$Physikalisches Institut, Universit{\"a}t Bonn,             
                   Bonn, Germany}                                             
\centerline{$^{22}$Physikalisches Institut, Universit{\"a}t Freiburg,         
                   Freiburg, Germany}                                         
\centerline{$^{23}$Institut f{\"u}r Physik, Universit{\"a}t Mainz,            
                   Mainz, Germany}                                            
\centerline{$^{24}$Ludwig-Maximilians-Universit{\"a}t M{\"u}nchen,            
                   M{\"u}nchen, Germany}                                      
\centerline{$^{25}$Fachbereich Physik, University of Wuppertal,               
                   Wuppertal, Germany}                                        
\centerline{$^{26}$Panjab University, Chandigarh, India}                      
\centerline{$^{27}$Delhi University, Delhi, India}                            
\centerline{$^{28}$Tata Institute of Fundamental Research, Mumbai, India}     
\centerline{$^{29}$University College Dublin, Dublin, Ireland}                
\centerline{$^{30}$Korea Detector Laboratory, Korea University,               
                   Seoul, Korea}                                              
\centerline{$^{31}$SungKyunKwan University, Suwon, Korea}                     
\centerline{$^{32}$CINVESTAV, Mexico City, Mexico}                            
\centerline{$^{33}$FOM-Institute NIKHEF and University of                     
                   Amsterdam/NIKHEF, Amsterdam, The Netherlands}              
\centerline{$^{34}$Radboud University Nijmegen/NIKHEF, Nijmegen, The          
                  Netherlands}                                                
\centerline{$^{35}$Joint Institute for Nuclear Research, Dubna, Russia}       
\centerline{$^{36}$Institute for Theoretical and Experimental Physics,        
                   Moscow, Russia}                                            
\centerline{$^{37}$Moscow State University, Moscow, Russia}                   
\centerline{$^{38}$Institute for High Energy Physics, Protvino, Russia}       
\centerline{$^{39}$Petersburg Nuclear Physics Institute,                      
                   St. Petersburg, Russia}                                    
\centerline{$^{40}$Lund University, Lund, Sweden, Royal Institute of          
                   Technology and Stockholm University, Stockholm,            
                   Sweden, and}                                               
\centerline{Uppsala University, Uppsala, Sweden}                              
\centerline{$^{41}$Physik Institut der Universit{\"a}t Z{\"u}rich,            
                   Z{\"u}rich, Switzerland}                                   
\centerline{$^{42}$Lancaster University, Lancaster, United Kingdom}           
\centerline{$^{43}$Imperial College, London, United Kingdom}                  
\centerline{$^{44}$University of Manchester, Manchester, United Kingdom}      
\centerline{$^{45}$University of Arizona, Tucson, Arizona 85721, USA}         
\centerline{$^{46}$Lawrence Berkeley National Laboratory and University of    
                   California, Berkeley, California 94720, USA}               
\centerline{$^{47}$California State University, Fresno, California 93740, USA}
\centerline{$^{48}$University of California, Riverside, California 92521, USA}
\centerline{$^{49}$Florida State University, Tallahassee, Florida 32306, USA} 
\centerline{$^{50}$Fermi National Accelerator Laboratory,                     
            Batavia, Illinois 60510, USA}                                     
\centerline{$^{51}$University of Illinois at Chicago,                         
            Chicago, Illinois 60607, USA}                                     
\centerline{$^{52}$Northern Illinois University, DeKalb, Illinois 60115, USA} 
\centerline{$^{53}$Northwestern University, Evanston, Illinois 60208, USA}    
\centerline{$^{54}$Indiana University, Bloomington, Indiana 47405, USA}       
\centerline{$^{55}$University of Notre Dame, Notre Dame, Indiana 46556, USA}  
\centerline{$^{56}$Purdue University Calumet, Hammond, Indiana 46323, USA}    
\centerline{$^{57}$Iowa State University, Ames, Iowa 50011, USA}              
\centerline{$^{58}$University of Kansas, Lawrence, Kansas 66045, USA}         
\centerline{$^{59}$Kansas State University, Manhattan, Kansas 66506, USA}     
\centerline{$^{60}$Louisiana Tech University, Ruston, Louisiana 71272, USA}   
\centerline{$^{61}$University of Maryland, College Park, Maryland 20742, USA} 
\centerline{$^{62}$Boston University, Boston, Massachusetts 02215, USA}       
\centerline{$^{63}$Northeastern University, Boston, Massachusetts 02115, USA} 
\centerline{$^{64}$University of Michigan, Ann Arbor, Michigan 48109, USA}    
\centerline{$^{65}$Michigan State University,                                 
            East Lansing, Michigan 48824, USA}                                
\centerline{$^{66}$University of Mississippi,                                 
            University, Mississippi 38677, USA}                               
\centerline{$^{67}$University of Nebraska, Lincoln, Nebraska 68588, USA}      
\centerline{$^{68}$Princeton University, Princeton, New Jersey 08544, USA}    
\centerline{$^{69}$State University of New York, Buffalo, New York 14260, USA}
\centerline{$^{70}$Columbia University, New York, New York 10027, USA}        
\centerline{$^{71}$University of Rochester, Rochester, New York 14627, USA}   
\centerline{$^{72}$State University of New York,                              
            Stony Brook, New York 11794, USA}                                 
\centerline{$^{73}$Brookhaven National Laboratory, Upton, New York 11973, USA}
\centerline{$^{74}$Langston University, Langston, Oklahoma 73050, USA}        
\centerline{$^{75}$University of Oklahoma, Norman, Oklahoma 73019, USA}       
\centerline{$^{76}$Oklahoma State University, Stillwater, Oklahoma 74078, USA}
\centerline{$^{77}$Brown University, Providence, Rhode Island 02912, USA}     
\centerline{$^{78}$University of Texas, Arlington, Texas 76019, USA}          
\centerline{$^{79}$Southern Methodist University, Dallas, Texas 75275, USA}   
\centerline{$^{80}$Rice University, Houston, Texas 77005, USA}                
\centerline{$^{81}$University of Virginia, Charlottesville,                   
            Virginia 22901, USA}                                              
\centerline{$^{82}$University of Washington, Seattle, Washington 98195, USA}  
}                                                                             

           
\begin{abstract}

From an analysis of the decay \bsdec\
we obtain the width difference between the light  and  
heavy  mass eigenstates,
$\Delta \Gamma \equiv (\Gamma_L - \Gamma_H)$
=  0.17 $\pm$ 0.09 (stat) $\pm 0.02 $ (syst) ps$^{-1}$ and 
the CP-violating phase  
$\phi _{s} = -0.79 \pm 0.56$ (stat) $^{+0.14}_{-0.01}$ (syst).
Under the hypothesis of  no CP violation ($\phi _{s} \equiv 0$), we obtain
$1/\overline \Gamma = \overline \tau(B^0_s)$ =1.52 $\pm 0.08 ^{+0.01}_{-0.03}$  ps and
$\Delta \Gamma =0.12 ^{+0.08} _{-0.10} \pm 0.02$ ps$^{-1}$.
The data sample corresponds to an integrated luminosity of 1.1 fb$^{-1}$
accumulated with the D0 detector at the Tevatron.

\end{abstract}


\maketitle

\newpage

In the standard model (SM), the light ($L$) and heavy ($H$) eigenstates 
of the mixed $B_s^0$ system  are expected
to have a sizeable 
mass and decay width difference, $\Delta M \equiv M_H - M_L$ and
$\Delta \Gamma \equiv \Gamma_L - \Gamma_H$. 
The CP-violating phase, defined as the the relative phase of the 
off-diagonal elements of the mass and decay 
matrices in the $B_s^0$ - $\overline B_s^0$ basis,  
 is predicted to be small. Thus, 
to a good approximation the two mass eigenstates are expected to be CP
eigenstates.  
New phenomena may alter the CP-violating mixing 
phase  $\phi _{s}$, 
leading to a reduction of the observed $\Delta \Gamma $
compared to the SM prediction~\cite{DFN2001}  $\Delta \Gamma_{SM} $:
$\Delta \Gamma$ =  $\Delta \Gamma_{SM} \times \cos\phi _{s}$.
While the mass difference
has recently been measured to  high precision~\cite{dmsd0,dms}, 
the CP-violating  phase remains unknown.

The decay \bsdec, proceeding through the quark process
$b\rightarrow c \bar c s$, 
gives rise to both CP-even and CP-odd final states.
It is possible to
separate the two CP components of the decay  \bsdec,
and thus  to measure the
lifetime difference, through a  study of
the time-dependent  angular distribution of the decay products of the 
$J/\psi$ and $\phi$ mesons.
Moreover,  with a  sizeable lifetime difference, there is 
sensitivity to the mixing phase 
through the interference
terms between the CP-even and CP-odd waves.

In Ref.~\cite{prl} we presented an analysis 
of  the decay chain \bsdec, $J/\psi \rightarrow \mu ^+ \mu ^-$,
$\phi \rightarrow K^+ K^-$ based on the first $\approx$450 pb$^{-1}$ 
of $p \bar p$ data at a center-of-mas energy of 1.96 TeV
collected  with the  D0 detector~\cite{run2det}.
In that analysis,
we extracted three parameters characterizing the \bs\ system and its
decay \bsdec: the average lifetime, 
$\overline \tau =1/\overline \Gamma$, where
$\overline{\Gamma}\equiv(\Gamma_H+\Gamma_L)/2$;
$\Delta \Gamma/\overline \Gamma$;
 and the  relative rate of the
decay to the CP-odd states at time zero.
Here we present new results, based on a two-fold
increase in statistics. 
In addition to  $\overline \tau$ and $\Delta \Gamma$, 
we extract for  the first time the
CP-violating phase $\phi _{s}$.
We also measure the magnitudes of the
decay amplitudes, and their relative phases. 

The data,  collected between June 2002 and January 2006,
correspond to an integrated luminosity of 1.1 fb$^{-1}$.
The selected 
events   include two reconstructed muons of  opposite charge,
with a transverse momentum
greater than 1.5 GeV and pseudorapidity  $|\eta| < 2$. 
 Each muon is required to be detected  as a track segment 
in at least one of the three layers of the muon system, and to 
be matched to a central track.
One muon is required to  have segments both inside and outside 
the toroid magnet.
We require the events to satisfy a muon trigger that does not include a cut
on the impact parameter.

\begin{figure}
\begin{center}\includegraphics[%
  width=8.0cm,
  height=8.5cm,
  keepaspectratio,
  trim=0 40 0 0
  ]{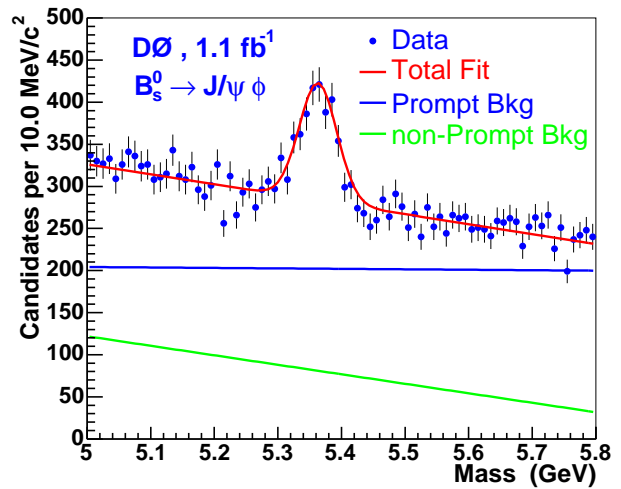}\end{center}

\caption{\label{fig:mass}
The invariant mass distribution of the ($J/\psi,\phi$) system for 
 \bs\ candidates. The curves are projections of the maximum likelihood fit (see text). 
}
\end{figure}

To select the $B_{s}^0$ candidate sample, we set the
minimum values of  momenta in the transverse plane for $B_{s}^0$, 
$\phi$, and $K$ meson candidates  at
6.0 GeV,    1.5 GeV, and 0.7 GeV, respectively.
$J/\psi$ candidates are accepted if the invariant mass of the muon pair
is in the range 2.9 -- 3.3 GeV.
For events in the central 
rapidity region (an event is considered to be central if the higher $p_T$
muon has $|\eta_{\mu 1}|<1$), 
we require the
transverse momentum of the $J/\psi$ meson to exceed 4 GeV. 
Successful candidates are constrained to the
world average mass of the $J/\psi$   meson~\cite{PDG}.
Decay products of the $\phi$ candidates are required 
to  satisfy a  fit to a common vertex and to have an  
invariant mass in the 
range 1.01 -- 1.03 GeV.
We require the ($J/\psi,\phi$) 
pair to be consistent with coming from a common vertex, and to have 
an invariant mass in the range 5.0 -- 5.8 GeV. 
In the case of multiple $\phi$ meson candidates, we
select the one with the highest transverse momentum.
Monte Carlo (MC) studies show that the $p_T$ spectrum of the $\phi$ mesons
coming from $B_s^0$ decay is harder than the spectrum of a pair
of random tracks from  hadronization.  
We define the signed  decay length of a \bs\ meson $L^B_{xy}$ 
as the vector pointing
from the primary vertex to the decay vertex projected on the
\bs\ transverse momentum.
To reconstruct the primary vertex, 
we select   tracks with   $p_{T}>$ 0.3 GeV
that are not used as decay products of the $B_s^0$ candidate,
and apply a constraint to the average beam spot position.
The  proper decay length, $ct$, is  
defined by the relation
$ct = L^B_{xy}\cdot M_{B^0_s}/p_T$ where $M_{B^0_s}$ is 
the measured mass of the $B_s^0$ candidate. 
The  distribution of the proper decay length  uncertainty $\sigma(ct)$  of \bs\ mesons 
 peaks around 25 $\mu$m. 
We accept events with $\sigma(ct) <60$  $\mu$m.
The invariant mass distribution of the accepted 23343 candidates  is
shown in Fig.~\ref{fig:mass}. The curves are   
projections of the maximum likelihood fit, described below.
The fit assigns  1039$\pm$45 (stat) events to the \bs\ decay.

We perform a simultaneous
 unbinned maximum likelihood fit to the proper decay length, 
three decay angles, and mass. 
The likelihood function ${\cal L}$ is given by:
\begin{eqnarray}
{\cal L} & = & \prod^{N}_{i=1}[ f_{sig}{\cal F}^i_{sig} + 
(1-f_{sig}){\cal F}^i_{bck}],
\end{eqnarray}
where $N$  is the total number of events, 
and $f_{sig}$ is the fraction of signal in the sample.
The function  ${\cal F}^i_{sig}$  describes the distribution of the
signal in  mass, proper decay length, and the decay angles, and 
${\cal F}^i_{bck}$ is the product of the background mass, proper decay length, 
and angular  probability density functions.
Background is divided into two categories. 
``Prompt'' background is due to directly 
produced $J/\psi$ mesons accompanied by random tracks arising from 
hadronization.  This background is distinguished from ``non-prompt'' 
background, where the $J/\psi$ meson is a product of a $B$ hadron decay 
while the tracks forming the $\phi$ candidate emanate from a multibody 
decay of the same $B$ hadron or from hadronization.

The time evolution of the angular distribution of the products of
the decay of {\it  flavor untagged}  $B_s^0$ mesons, i.e., summed over
 $B^0_s$ and $\overline{B}^0_s$, expressed in terms of the linear 
polarization amplitudes $A_x$ and their relative phases $\delta_i$
is~\cite{DFN2001}: 

\begin{widetext}
$$
\frac{d^3 \Gamma (t)}
{d \cos \theta~d \varphi~d \cos \psi}
\propto ~2 |A_0(0)|^2  \ {\cal{T_{+}}}\  \cos^2\psi
(1 - \sin^2 \theta\cos^2 \varphi)
\hfill{ } 
+ \sin^2 \psi \{ |A_\parallel(0)|^2   \ {\cal{T_{+}}}\
(1 - \sin^2 \theta \sin^2 \varphi)
+ |A_\perp(0)|^2   \ {\cal{T_{-}}}\  \sin^2 \theta \} ~~~
$$
$$
+\frac{1}{\sqrt{2}}  \sin 2 \psi |A_0(0)|| A_\parallel(0)|
\cos(\delta_2-\delta_1)  \ {\cal{T_{+}}}\  \sin^2\theta  \sin 2 \varphi
$$
\beq\label{tripleuntagged}
+\biggl\{\frac{1}{\sqrt{2}}|A_0(0)|| A_\perp(0)|\cos\delta_2\sin2\psi
\sin2\theta\cos\varphi
$$
$$
-|A_\parallel(0)|| A_\perp(0)|\cos\delta_1\sin^2\psi
\sin2\theta\sin\varphi\biggr\}\frac{1}{2}\left(e^{-\Gamma_H t} -
  e^{-\Gamma_L t}\right) \sin\phi_s~.
\eeq
\end{widetext}
where  
 ${\cal{T_{+/-}}} = \frac{1}{2} \left( (1 \pm\cos\phi_s)e^{-\Gamma_{L}t}+(1 \mp\cos\phi_s)e^{-\Gamma_{H}t} \right)$.

In the coordinate system of the $J/\psi$ rest frame 
 (where the $\phi$ meson moves in the $x$ direction,
 the $z$  axis is perpendicular to 
the decay plane of $\phi \to K^+ K^-$, and $p_y(K^+)\geq 0$),
the transversity polar and azimuthal angles 
$(\theta, \varphi)$ describe the
direction of the $\mu^+$, and $\psi$ is 
the angle between   $\vec p(K^+)$ and  $-\vec{p}(J/\psi)$ 
 in the $\phi$ rest frame.

We model the acceptance in the three angles by fits using 
polynomial functions, with parameters determined using Monte Carlo simulations.
We have used  the {\sl SVV\_HELAMP} model
in the {\sc {EvtGen}} generator \cite{evtgen}, interfaced to the {\sc Pythia}
 program \cite{pythia}. Simulated events 
were reweighted to match the kinematic distributions observed in
the data.

The lifetime distribution shape  of the  background is described as a sum of
a prompt component, simulated as 
a Gaussian function centered at zero, and a non-prompt component,
simulated as a superposition of one exponential for the negative $ct$ 
region and two exponentials
for the positive $ct$ region, with free slopes and normalization.
The mass distributions of the backgrounds are parametrized by 
first-order
polynomials.
 The distributions in the transversity polar and azimuthal angles 
are parametrized 
as $1+X_{2x}\cos^2\theta + X_{4x}\cos^4\theta$ and
$1+Y_{1x} \cos(2\varphi) +Y_{2x} \cos^2(2\varphi)$, respectively.
For the background dependence on the angle $\psi$, we use the function
$1+Z_{2x} \cos^2(\psi)$.
We also allow for a background term analogous to the interference term
of the CP-even waves,
with one free coefficient. For each of the 
above background functions  we use two separate sets of parameters 
for  the prompt and non-prompt  components.

\begin{figure}[htb]
\begin{center}\includegraphics[%
  width=8.0cm,
  height=8.5cm,
  keepaspectratio,
  trim=0 40 0 0
  ]{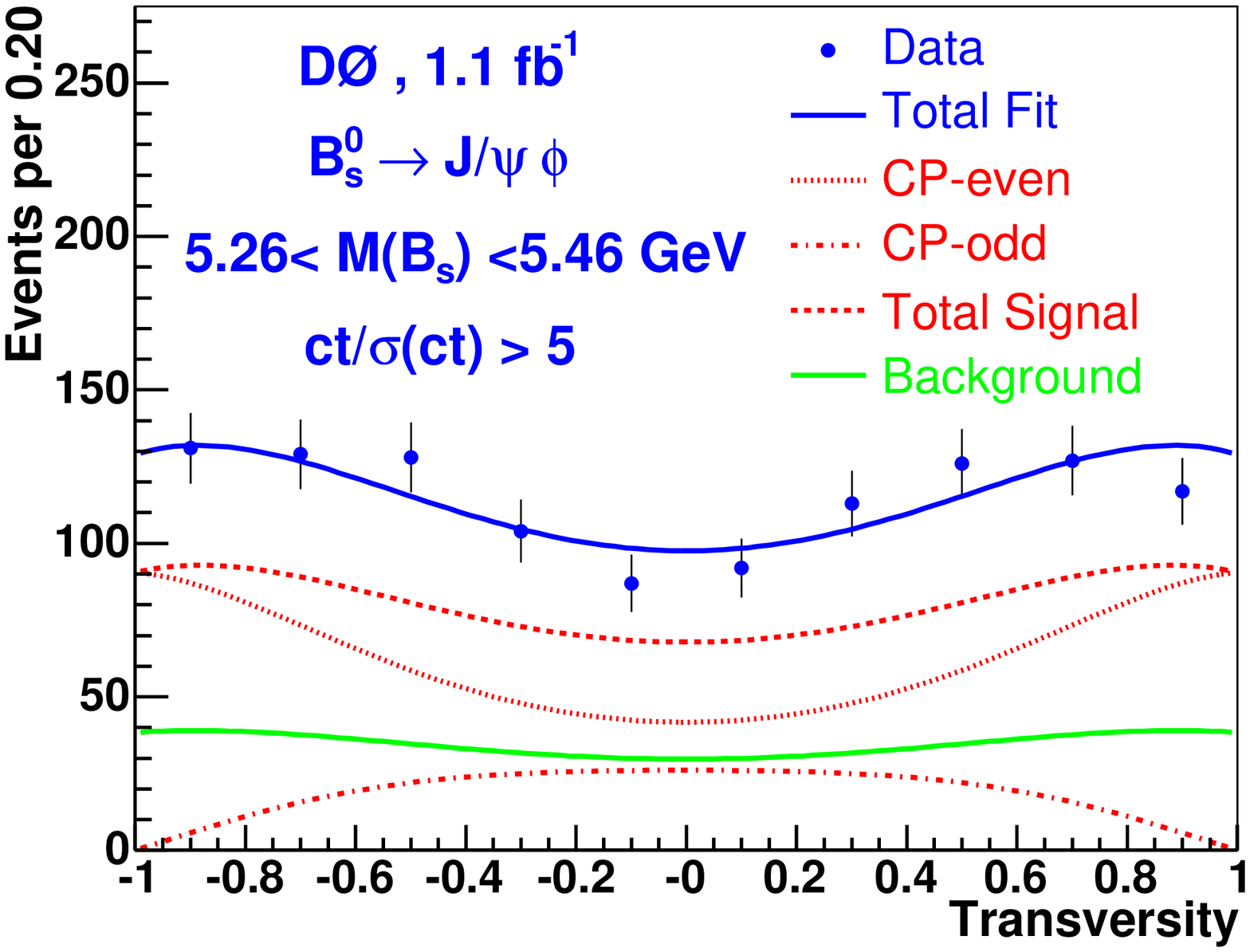}\end{center}

\caption{
The transversity polar angle distribution for 
the signal-enhanced subsample: $ct/\sigma(ct)>5$ and signal mass
range.
The curves show:  the signal contribution,  dotted (red); 
the background, light solid (green); and total, solid (blue) [color online].
}
\label{fig:bs_trans_fit_sig}
\end{figure}

\begin{figure}[htb]
\begin{center}\includegraphics[%
  width=8.0cm,
  height=8.5cm,
  keepaspectratio,
  trim=0 40 0 0
  ]{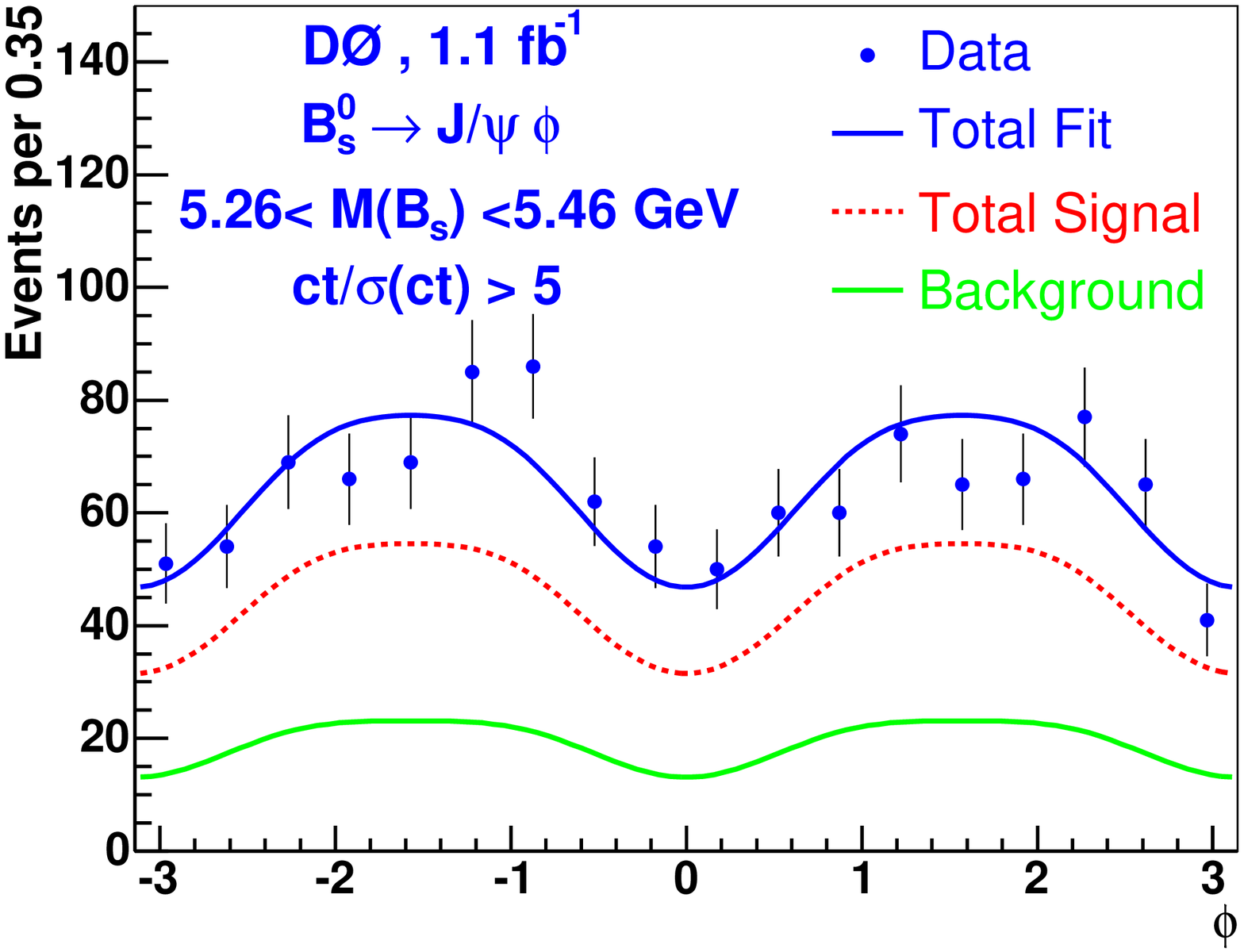}\end{center}

\caption{
The transversity asimuthal angle distribution for 
the signal-enhanced subsample:   $ct/\sigma(ct)>5$ and signal mass
range.
The curves show:  the signal contribution,  dotted (red); 
the background, light solid (green); and total, solid (blue) [color online].
}
\label{fig:bs_phia_fit_sig}
\end{figure}

\begin{figure}[htb]
\begin{center}\includegraphics[%
  width=8.0cm,
  height=8.5cm,
  keepaspectratio,
  trim=0 40 0 0
  ]{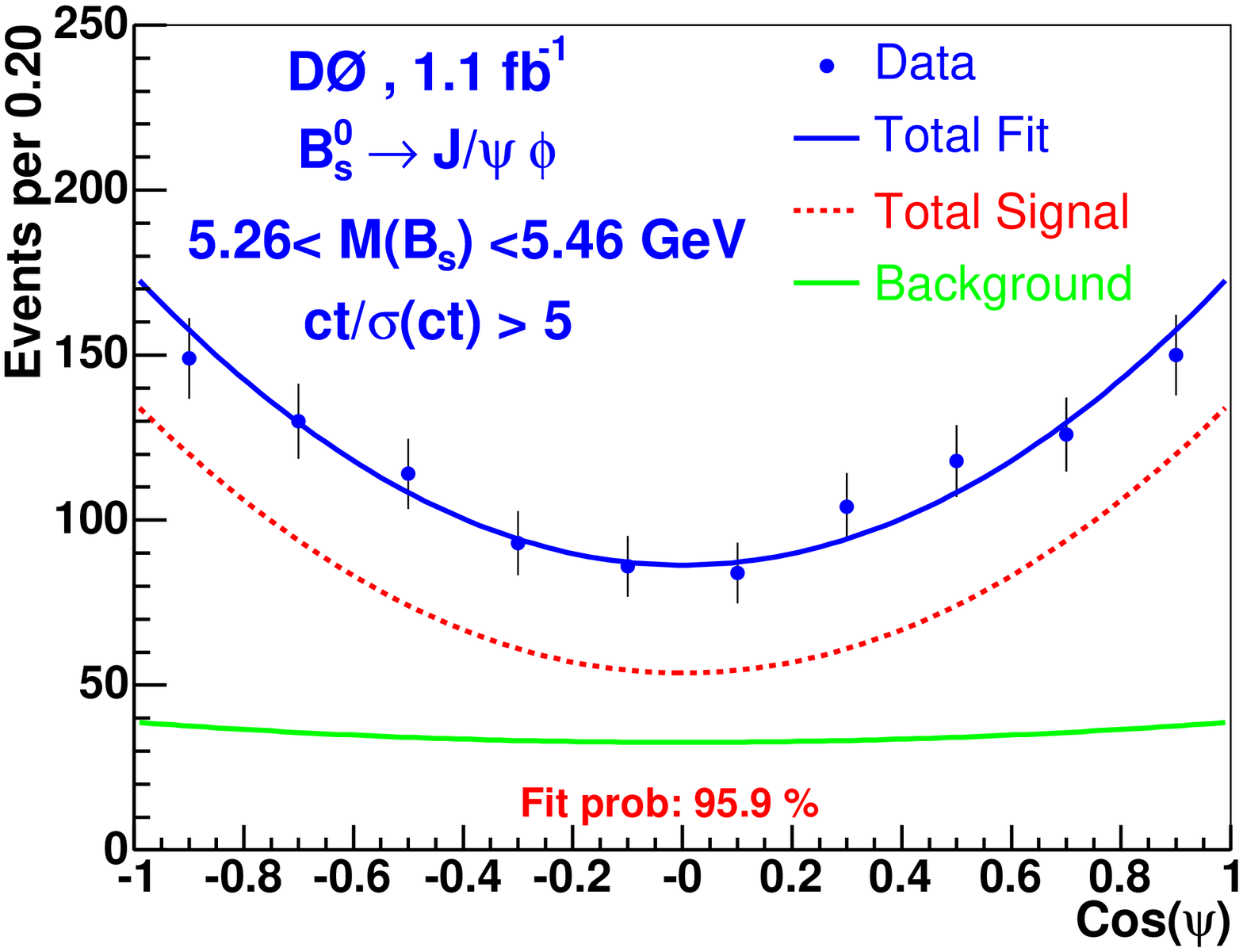}\end{center}

\caption{
The $\psi$  angle distribution for 
the signal-enhanced subsample:  $ct/\sigma(ct)>5$ and signal mass
range.
The curves show:  the signal contribution,  dotted (red); 
the background, light solid (green); and total, solid (blue) [color online].
}
\label{fig:bs_psia_fit_sig}
\end{figure}

\begin{figure}[htb]
\begin{center}\includegraphics[%
  width=8.0cm,
  height=8.5cm,
  keepaspectratio,
  trim=0 40 0 0
  ]{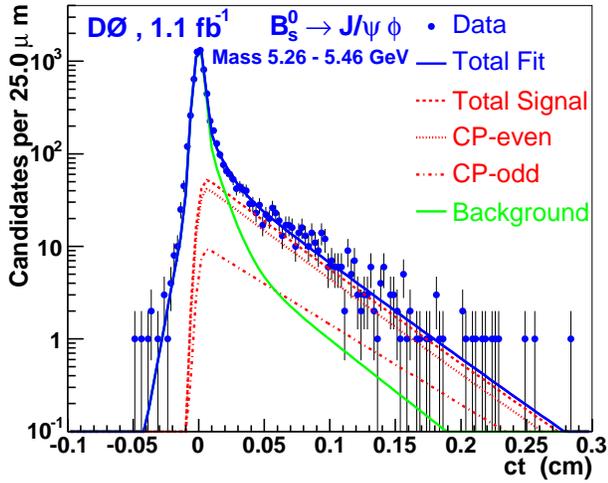}\end{center}
\caption{The proper decay length, $ct$,  of the \bs\ candidates
in the signal mass region. 
The curves show:  the signal contribution,  dashed (red);
the CP-even (dotted) and CP-odd (dashed-dotted) contributions 
of the signal, 
the background, light solid(green); and total, solid (blue) [color online].}
\label{fig:bs_lifetime_sig}
\end{figure}

\begin{figure}[h!tb]
\begin{center}\includegraphics[%
  width=8.0cm,
  height=8.5cm,
  keepaspectratio,
  trim=0 40 0 0
  ]{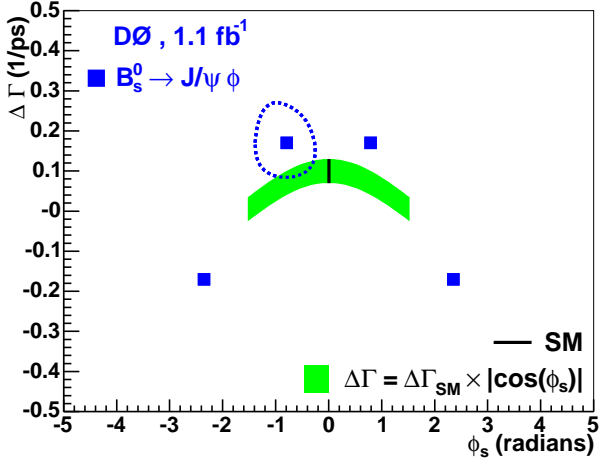}\end{center}
\caption{
The  $\Delta\ln({\cal L}) = 0.5$ contour (error elipse)
in the plane  ($\Delta \Gamma $, $\phi _{s}$) for the
fit to the \bsdec\ data.
Also shown is the 
band representing the relation
$\Delta \Gamma$ =  
$\Delta \Gamma_{SM} \times |(\cos(\phi _{s})|$, with  
$\Delta \Gamma_{SM} = 0.10 \pm 0.03$ ps$^{-1}$~\cite{lenz}.
The 4-fold ambiguity is discussed in the text.
}
\label{fig:bs_contdgvsdphi}
\end{figure}

Our results for the hypothesis of CP conservation 
and for the case of free $\phi _{s}$,
are presented in Table~\ref{bs_rperp}.
For the CP-violating phase, which has a four-fold ambuguity discussed below,
the fit value closest to the SM prediction of  $-0.03$~\cite{DFN2001} is 
$\phi _{s} = -0.79 \pm 0.56$.
Figures~\ref{fig:bs_trans_fit_sig} -- \ref{fig:bs_lifetime_sig} show
the fit projections on the angular distributions and the proper decay length.
Figure~\ref{fig:bs_contdgvsdphi} shows the  
$\Delta\ln({\cal L}) = 0.5$ error ellipse contour 
(corresponding to the confidence level of 39\%)
in the plane  ($\Delta \Gamma $, $\phi _{s}$).
As seen from Eq.~\ref{tripleuntagged},
the sign of  $\sin\phi _{s}$ is reversed with the simultaneous reversal
of the signs of $\cos\delta_1$ and  $\cos\delta_2$.
For the case  $\cos\delta_1<0$ and  $\cos\delta_2>0$,
our measurement correlates two possible solutions for $\phi _{s}$ with the
two signs of $\Delta \Gamma$: $\phi _{s}=-0.79\pm$0.56,
$\Delta \Gamma > 0$, and  $\phi _{s} = 2.35\pm$0.56, $\Delta \Gamma < 0$.
For the case  $\cos\delta_1>0$ and  $\cos\delta_2<0$ the two solutions
are  $\phi _{s} = 0.79\pm$0.56,
$\Delta \Gamma > 0$, and  $\phi _{s} = -2.35\pm$0.56, $\Delta \Gamma < 0$.

\begin{table}[h!tb]
\renewcommand{\arraystretch}{1.3}
    \begin{center}
\caption {Maximum likelihood fit results.
Sign ambiguities are discussed in the text.
}
.

\begin{tabular}{lccc}
 \hline
Observable &  CP conserved & free $\phi _{s}$  \\ \hline 
${\textstyle \Delta\Gamma \ (\rm{ps}^{-1})}$   & $0.12^{+0.08}_{-0.10} $ & $0.17^{+0.09}_{-0.09}$ & \tabularnewline
${\textstyle \frac{1}{\overline{\Gamma}} = \overline{\tau} \ ({\rm{ps}})}$ & $1.52^{+0.08}_{-0.08}$ & $1.49 \pm 0.08$ \\ 
${\textstyle \phi _{s} }$              & $\equiv 0$ & $-0.79\pm$0.56 \\
      
${\textstyle |A_0(0)|^2-|A_\parallel(0)|^2 }$ & 0.38$\pm$0.05 & 0.37$\pm$0.06 \\ 
${\textstyle A_{\perp}(0) }$ & 0.45$\pm$0.05 & 0.46$\pm$0.06 \\
${\textstyle \delta_1-\delta_2 }$ & 2.6$\pm$0.4 & 2.6$\pm$0.4 \\ %
${\textstyle \delta_1 }$ & -- & 3.3$\pm$1.0 \tabularnewline 
 ${\textstyle \delta_2 }$  & -- & 0.7$\pm$1.1 \tabularnewline \hline
\label{bs_rperp}
\end{tabular}
\end{center}
\end{table}

We perform a  test using  pseudo-experiments with similar
statistical sensitivity, generated with the same parameters
as obtained in this analysis  under the condition of no CP violation.
When fits allowing for CP violation are performed, $\approx$ 50\% 
of the experiments  have a fitted $\cos(\phi _{s})$ less than the measured value.
About 80\% of experiments have the statistical uncertainty of
 $\phi _{s}$ greater than that for data.

We verify the  procedure by performing  fits on MC
samples passed through the full chain of  detector 
simulation, event reconstruction, and maximum likelihood fitting.
We assign systematic uncertainties due to the statistical
precision of this procedure test.
We also repeat the fits to the data with the parameters describing 
the acceptance  varied by $\pm 1\sigma$.

Uncertainties from the data processing
reflect the stability of the results with respect to different versions
of the track and vertex 
reconstruction algorithms. 
The ``interference'' term in the background model accounts for
the collective effect of various physics processes. However, its
presence may be partially due to the detector
acceptance effects. Therefore, we interpret the difference between fits with 
and without this term as a systematic uncertainty associated with
 the background model.
Effects of the imperfect detector alignment are estimated
using a modified geometry of the the silicon microstrip tracker,
with silicon sensors moved within the known uncertainty.  
The effects of systematic uncertainties are listed in Table \ref{syst}.

\begin{table}[h!tb]
\caption {Sources of  systematic uncertainty in the results of the
analysis of the decay \bsdec. 
}
\begin{tabular}{ccccccc}
\hline
Source    &  $c\tau(B_s^0)$  & $\Delta \Gamma$ & $\rperp$ & 
 \ \ $\phi_s$ \tabularnewline
&  $\mu$m & ps$^{-1}$ &&&&\tabularnewline
\hline
Procedure test                        &$\pm$2.0    &$\pm$0.02  &$\pm 0.01$ & -- \tabularnewline
Acceptance    &$\pm 0.5$  & $\pm 0.001$ &$\pm 0.003$ & \ \  $\pm 0.01$\tabularnewline
Reco. algorithm& $-8.0$,$+1.3$  & +0.001 & $\pm 0.01$ & $-0.01$ \tabularnewline
Background model& +1.0  & +0.01 & $-0.01$ & +0.14 \tabularnewline
Alignment     &$\pm$2.0  & --  & --   & \ \ -- \tabularnewline
Total      &$-8.8,+3.3$  &$\pm 0.02$  &$\pm 0.02$ & \ \ $-0.01,+0.14$ \tabularnewline
\hline
\end{tabular}
\label{syst}
\end{table}

From a fit to the CP-conserving time-dependent 
angular distribution 
of the untagged decay \bsdec, 
we obtain the average lifetime of the $B_s^0$ system,  
$\overline \tau(B^0_s) =1.52 \pm 0.08$ (stat) $^{+0.01}_{-0.03}$ (syst) ps
and the  width difference between the two mass eigenstates, 
$\Delta \Gamma =0.12 ^{+0.08}_{-0.10}$ (stat) $\pm 0.02$ (syst) ps$^{-1}$.

Allowing for CP violation in $B_s^0$ mixing,
we provide the first direct constraint on the CP-violating phase, 
$\phi _{s} = -0.79 \pm 0.56$ (stat) $^{+0.14}_{-0.01}$ (syst).

%
We thank U. Nierste for useful discussions.
We thank the staffs at Fermilab and collaborating institutions, 
and acknowledge support from the 
DOE and NSF (USA);
CEA and CNRS/IN2P3 (France);
FASI, Rosatom and RFBR (Russia);
CAPES, CNPq, FAPERJ, FAPESP and FUNDUNESP (Brazil);
DAE and DST (India);
Colciencias (Colombia);
CONACyT (Mexico);
KRF and KOSEF (Korea);
CONICET and UBACyT (Argentina);
FOM (The Netherlands);
PPARC (United Kingdom);
MSMT (Czech Republic);
CRC Program, CFI, NSERC and WestGrid Project (Canada);
BMBF and DFG (Germany);
SFI (Ireland);
The Swedish Research Council (Sweden);
Research Corporation;
Alexander von Humboldt Foundation;
and the Marie Curie Program.
%

\end{document}